\documentclass[preprint,aps,prd]{revtex4-2}
\usepackage{lipsum}
\usepackage{graphicx}
\usepackage{subcaption} 
\usepackage{ragged2e} 
\usepackage{dcolumn}
\usepackage{makecell} 
\usepackage{bm}
\usepackage{setspace}
\usepackage{xcolor, soul} 
\usepackage{amsmath}
\usepackage{comment}
\usepackage{hyperref}
\hypersetup{
    colorlinks,
    linkcolor=black,
    citecolor=black,
    urlcolor=black
}

\begin{document}

\title{Cascades in the Kinetic Equation for the Majda-McLaughlin-Tabak model}

\author{
Gregorio Tibone$^1$, Giorgio Krstulovic$^1$ and Miguel Onorato$^{2,3}$}

\affiliation{%
 $^1$ Universit\'e C\^ote d’Azur, CNRS, Institut de Physique de Nice (INPHYNI), 17 rue Julien Laupr\^etre, 06200 Nice, France \\
$^2$ Dipartimento di Fisica, Università di Torino, Via P. Giuria 1, 10125, Torino, Italy \\
$^3$ INFN, Sezione di Torino, Via P. Giuria 1, Torino, 10125, Italy}

\begin{abstract}
\setstretch{1.2}
The Majda-McLaughlin-Tabak (MMT) family of models has proven to be an efficient ground for benchmarking wave turbulence theory, thanks to the low computational cost required to test theoretical ideas and the possibility of tuning nonlinearity and dispersive properties of the equations. Here, we study numerically the wave kinetic equation (WKE) associated with the MMT model and perform simulations to study turbulent cascades. We confirm numerically the predictions of wave turbulence theory, both in the parameter space region where the wave kinetic equation was proven to be well posed and outside of it. We also observe a new stable stationary state in a region where no cascade solutions are expected, a region that, to the best of our knowledge, has not been explored before. 
Moreover, following recent work, we study next-to-leading-order corrections to the wave kinetic
equation; we uncover incurable divergences in the one-dimensional MMT model and, more generally,
in higher-dimensional systems with concave power-law dispersion relations.
\end{abstract}

\maketitle

\newpage

\setstretch{1.1}

\section{Introduction}

While the kinetic theory for a gas of interacting particles was introduced at the end of the  19th century  by Boltzmann \cite{boltzmann1970weitere}, the statistical theory of interacting dispersive waves emerged much later. Aside from Peierls’ pioneering work on phonons \cite{peierls1929kinetischen}, significant progress was not achieved until the 1960s, when the theory was systematically developed across several areas of physics. The seminal contributions of Hasselmann \cite{Hasselmann1962} and Zakharov \cite{Zakharov1965} laid the foundations of the modern framework. 
Both authors, starting from surface gravity wave dynamical equations, derived in a formal way the wave kinetic equation, an evolution equation for the wave action spectral density, which is related to the wave amplitude. 
In addition to the thermal equilibrium stationary state, known as the Rayleigh–Jeans spectrum, the Wave Kinetic Equation also admits solutions characterized by constant fluxes of conserved quantities. These non-equilibrium steady-state solutions are known as Kolmogorov–Zakharov (KZ) spectra \cite{zakharov2025kolmogorov,Nazarenko2011}. Although a formal derivation of the Wave Kinetic Equation is relatively straightforward, establishing its validity up to arbitrary time scales is far from trivial. To date, such rigorous results have been obtained only for a limited class of specific nonlinear dispersive wave systems \cite{deng2023full}. 
The derivation relies on perturbation theory, subtle limiting procedures (such as the large-box and weak-nonlinearity limits), and the assumption that randomness in the wave field persists over time. These requirements are not always straightforward to realize in natural environment or in numerical simulations of the underlying dynamical equations. Therefore, establishing the applicability of the Wave Kinetic Equation for an a general waves system is highly non-trivial.

In 1997, Majda, McLaughlin and Tabak \cite{majda1997one} questioned the validity of the Wave Kinetic Equation by introducing a seemingly simple family of one-dimensional nonlinear dispersive wave equations, known as the MMT model. After comparing direct numerical simulations of the underlying dynamical system with the predictions of the kinetic equation, they claimed that the predictions of weak turbulence theory fail and yield a much flatter spectrum than the steeper one observed in the numerical statistical steady state. 
Such claim originated a large amount of literature trying to explain such behavior (\cite{ZAKHAROV2004,chibbaro2017weak,simonis2024transition,cai2001dispersive,hrabski2024verification,zakharov2001wave,pushkarev2013quasibreathers,du2023impact,rumpf2005weak} among others). The most recent simulations \cite{hrabski2024verification} support Zakharov’s original hypothesis \cite{zakharov2001wave} that the emergence of coherent structures under stronger nonlinear conditions leads to deviations from the spectral slopes predicted by Wave Turbulence theory. 

The MMT model has the following form:
\begin{equation}\label{eq:mmt}
i\partial_t=|\partial_x|^{\alpha}\psi+\sigma|\partial_x|^{\beta/4}(||\partial_x|^{\beta/4} \psi|^2\partial_x|^{\beta/4}\psi),\,
\end{equation}
where $\psi(x,t)$ is a complex wave field, $\beta$ and $\alpha$ are real parameters that characterize the nonlinear dispersive properties of the model, the operator $|\partial_x|^\alpha f(x)$ is defined in Fourier space as $|k|^\alpha \hat{f}(k)$, and $\sigma=\pm 1$ identifies the de-focusing and the focusing regime. For $\beta=0$ and $\alpha=2$, the system is equivalent to the Nonlinear Schr\"odinger equation, while for $\alpha=1/2$, the linear dispersion relation is equivalent to the one of surface gravity waves in infinite water depth.

Starting from equation (\ref{eq:mmt}) and following the procedures outlined in \cite{zakharov2025kolmogorov} or \cite{Nazarenko2011},  one can formally derive an evolution equation for the wave action spectral density, which leads to the Wave Kinetic Equation. However, from a mathematical standpoint, this derivation is not fully rigorous.
The only rigorous justification to date is provided in \cite{vassilev2025one}, where it is shown that the kinetic equation, for $0<\alpha<1$ remains valid up to a fraction of the kinetic time. The well-posedness of the Wave Kinetic equation associated to the MMT was  investigated in \cite{germain2025local}, where it was established that, for $\alpha=1/2$, the collision integral is bounded for $-1/2 \le \beta \le 0$. 
Formally, the Kinetic Equation for the MMT model admits KZ stationary solutions for $\alpha\in (0,1)$ and arbitrary $\beta$; however, true solutions are only those for which the collision integral converges (in the wave turbulence community this property is known as {\it locality}); therefore, the properties of the integrals in 0 and $\infty$ should be checked. This study has been presented in \cite{ZAKHAROV2004} and it was found that convergences is guaranteed if $\beta<3(2-\alpha)$. Moreover, because the KZ solutions are characterized by constant fluxes, one should also check for which values of $\beta$ and $\alpha$ the energy flux is positive (direct cascade) and  the wave action flux is negative (inverse cascade). The results obtained in \cite{ZAKHAROV2004} is  that the fluxes of wave action and energy simultaneously have the correct signs in the region of parameter $\beta<-3/2$ and $\beta>2\alpha-3/2$.
All these issues related to locality and signs of the fluxes  will be further discussed in the present paper. 

It is evident from the discussion above that Wave Turbulence theory is rich in detail and potentially, if one is interested in observing KZ solutions, the situation is more subtle than it initially appears.
One should keep in mind that one of the central assumptions underlying Wave Turbulence theory is the weakness of the nonlinear interactions. But what happens when this assumption is no longer satisfied? This question has recently been addressed from a theoretical perspective. In particular, Rosenhaus and Smolkin \cite{rosenhaus2024wave} derived a next-to-leading-order correction to the kinetic equation, going beyond the conventional weakly nonlinear approximation. Their work provides a systematic framework to quantify deviations from the standard theory and to assess the robustness of its predictions. In this framework, the ultraviolet and infrared divergences associated with KZ solutions, already delicate within the leading-order theory, have been further investigated in \cite{Rosenhaus:2023pdj,schubring2025interplay}. 

In this work, we focus on the MMT model, studying its non-equilibrium steady states obtained from its WKE in different regions of parameters.
In Section \ref{sec:kin_theory}, we briefly introduce the kinetic theory of the MMT model, highlighting the salient points related to stationary solutions and their fluxes behavior.   
In Section \ref{sec:num_sim}, we numerically simulate the WKE, finding good agreement with the predictions of WT theory for $\alpha=1/2$ and $\beta \geq -1/2$, even outside of the parameter range where the equation was proven to be well-posed. Then, looking at the region $\beta \leq -1/2$, where no KZ solution is expected, we uncover a new type of stationary state. 
In Section \ref{sec:beyond}, we discuss current approaches to describe wave kinetic dynamics beyond the leading order, in particular for the MMT model. We show that the next to the leading-order correction to the kinetic equation is highly divergent, and therefore breaks the convergence of the small non-linearity expansion. We then show that this strong divergence is generic in any dimension for all dispersive wave systems with $\alpha < 1$.
Conclusions are reported in Section \ref{sec:conclusions}.

\section{Kinetic Theory of the MMT model}\label{sec:kin_theory}
Following \cite{ZAKHAROV2004}, we move from a dynamical to a statistical description of the model under the usual assumptions of observing the late time dynamics, phase randomization along time evolution and weak nonlinearity. 
%
The resulting theory consists in a probability distribution for the normal variables $\psi_k(t)$, whose second moment is defined as $\langle \psi_k(t) \psi^*_{k'}(t) \rangle = n_k(t) \delta(k-k')$,  and is governed by the celebrated wave kinetic equation (\cite{zakharov2025kolmogorov}, \cite{Nazarenko2011})
\begin{gather}
    \frac{\partial}{\partial t}n_k = \mathcal{S}\{n_k\}\notag \\
    \mathcal{S}\{n_k\} = 4\pi \int_{-\infty}^\infty dk_1 dk_2 dk_3 |k k_1 k_2 k_3|^{\beta/2}n_k n_1 n_2 n_3 
    \left( \frac{1}{n_k} + \frac{1}{n_1} - \frac{1}{n_2} - \frac{1}{n_3}\right) \delta\left(\Delta k^{k1}_{23}\right)\delta\left(\Delta\omega^{k1}_{23}\right),
    \label{MMT_wke}
\end{gather} 
where for a generic function $f_k$ is a shorthand for $f(k)$ and $\Delta f^{ij}_{hl} = f_i + f_j - f_h - f_l$. Moreover, the time dependence of $n_k(t)$ is suppressed for readability. $ S\{n_k\}$ is usually referred to as the collisional integral. For a fixed external wavevector $k$, the Dirac delta functions in the collisional integral define a one dimensional manifold in the domain of integration; we shall refer to it as the resonant manifold.
In the following, the solutions considered will always be isotropic in 1D, i.e., $n_k = n_{-k}$. 

It is simple to prove that, during the time evolution, the linear energy $E = \int_{-\infty}^\infty  n_k \omega_k dk$, and wave action $N = \int_{-\infty}^\infty  n_k dk$ are conserved. Their corresponding local conservation laws are 
\begin{align}
    2 \frac{\partial(\omega_kn_k)}{\partial t}  + \frac{\partial P_k}{\partial k}  &= 0 
    \label{conservation energy}\\
    2\frac{\partial n_k }{\partial t} + \frac{\partial Q_k}{\partial k}  &= 0,
    \label{conservation wave action}
\end{align}
where $P_k$ and $Q_k$ are the energy and the wave action fluxes, respectively. They follow directly from the collisional integral. 
For this statistical theory, it is possible to define entropy in such a way that an H-theorem holds (\cite{zakharov2025kolmogorov}). Moreover, at its maximum, the stationary solution of eq.\eqref{MMT_wke} is the Rayleigh-Jeans (RJ) distribution, whose form is:
\begin{equation}
    n_k = \frac{T}{\omega_k + \mu},
    \label{RJ}
\end{equation}
with $T$ and $\mu$ usually called temperature and chemical potential, functions of $N$ and $E$.
Both the energy and wave action fluxes are null for the RJ solution. 

\subsection{Kolmogorov-Zakharov flux solutions}
Zakharov showed \cite{zakharov2025kolmogorov} that kinetic equations like \eqref{MMT_wke} possess other exact stationary solutions, with constant non-zero fluxes. Those are power laws and are called Kolmogorov-Zakharov (KZ) solutions. For the MMT model, they are of the form:  
\begin{align}
    n_k^P &= A k^{-\gamma_p} , \quad \gamma_p = 1 - \alpha/3 +  2\beta/3,
    \label{KZ P}\\ 
    n_k^Q &= A k^{-\gamma_q}, \quad \gamma_q = 1 + 2\beta/3.
    \label{KZ Q}
\end{align}
To be considered acceptable, those solutions must be local, meaning that the collisional integral and its derivative with respect to the solution exponent converge. It needs also to be stable, meaning that perturbations of the stationary state decrease with time. The reader is directed to  \cite{Balk2000} or \cite{Costa2026} for a more in-depth discussion of these topics. To the best of our knowledge, the stability of MMT solutions has not yet been demonstrated: however, the KZ solutions are expected to be stable from the numerous numerical simulations performed during the years.
Those two solutions correspond, respectively, to a constant energy flux and a constant wave action one. A standard argument, due to Fjortoft, suggests that, in the limit of an infinite inertial range, energy flows towards high $k$, small scale, and wave action towards low $k$, large scale. The region where no conserved quantities are introduced or dissipated, and thus where eq. \eqref{MMT_wke} holds, is called the inertial range.       
When substituting a generic power law $Ak^{-\nu}$ inside the collisional integral, thanks to homogeneity, the collisional integral can be re-expressed as 
\begin{equation}
    \mathcal{S}\{n_k\} = 8\pi A^3 k^{2 - 3\nu + 2\beta - \alpha} \mathcal{I}(\nu),
    \label{dimensionless_coll}
\end{equation}
where $\mathcal{I}(\nu)$ is the dimensionless part of the integral, given explicitly in Appendix \ref{section:Aappendix}.
Comparing eq. \eqref{conservation energy} and \eqref{conservation wave action} with \eqref{MMT_wke}, one can derive  
\begin{align}
    P_0 =& \frac{8\pi {A}^3}{3} \mathcal{I}'(\gamma_p) 
    \label{collisional flux energy}\\
    Q_0 =& \frac{8\pi {A}^3}{3} \mathcal{I}'(\gamma_q),
    \label{collisional flux wave action}
\end{align}
where the primed symbols denote a derivative with respect to the exponent $\nu$, and we have assumes that because stationary $\mathcal{I}(\gamma_p)=\mathcal{I}(\gamma_s)=0$. $P_0$ denotes the energy flux relative to solution \eqref{KZ P} and $Q_0$ denotes the same for $\eqref{KZ Q}$. 
Overall, we can rewrite solutions \eqref{KZ P} and \eqref{KZ Q} as 
\begin{align}
    n_k^P &= C_{KZ}^P {P_0}^{1/3} k^{-\gamma_p} , \quad C_{KZ}^P = \left(\frac{8\pi\mathcal{I}'(\gamma_p) }{3}\right)^{-1/3},
    \label{KZ P 2}\\ 
    n_k^Q &= C_{KZ}^Q {Q_0}^{1/3} k^{-\gamma_q} , \quad C_{KZ}^Q = \left(\frac{8\pi\mathcal{I}'(\gamma_q) }{3}\right)^{-1/3},
    \label{KZ Q 2}
\end{align}
where the two $C_{KZ}$ are called Kolmogorov-Zakharov constants. 
For completeness, we present in table \ref{tab:KZ_constants} the values of the KZ constants for the values of $\beta$ that we have explored numerically.
\begin{table}[h!]
    \centering
    \setlength{\tabcolsep}{10pt}
    \renewcommand{\arraystretch}{1.3}
    \begin{tabular}{|c|c|c|}
        \hline
        $\beta$ & \makecell{$C_{KZ}^P$} & \makecell{$C_{KZ}^Q$} \\
        \hline
        $0$ & 0.270 & -0.370 \\
        $1$ & 0.157 & -0.202 \\
        $- 0.51$ & 0.453 & 1.904 \\
        \hline
    \end{tabular}
    \caption{Values of the KZ constants for $\alpha = 1/2$ and values of $\beta$ explored. Notice how in the first two rows the signs of the two constants are different, whereas in the third one both are positive.}
    \label{tab:KZ_constants}
\end{table}

\begin{figure}[h!]
    \centering
    \includegraphics[width=0.6\linewidth]{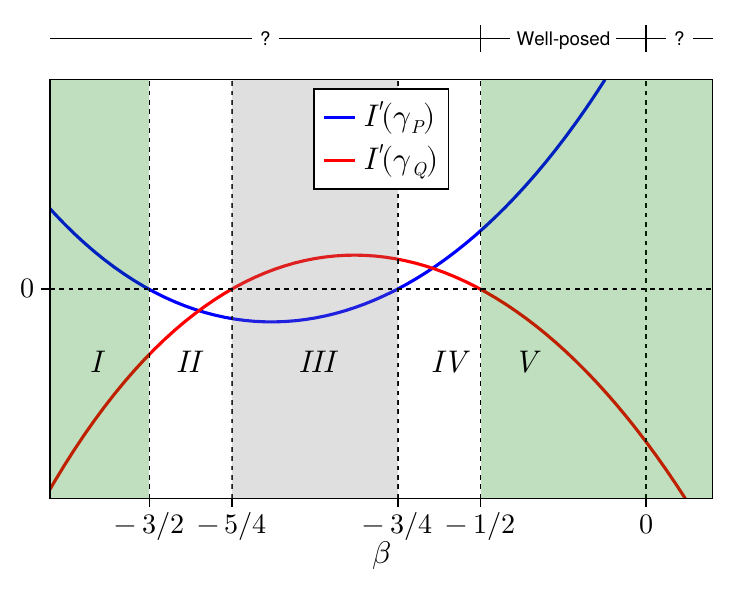}
    \\
    \begin{tabular}{|c|c|c|c|c|}
\hline
 & $Q_0$ & $P_0$ & \makecell{Energy \\ cascade} & \makecell{Wave action \\ cascade}\\
\hline
I & $<0$ & $>0$  & KZ & KZ \\
II & $<0$ & $<0$  & warm & KZ \\
III & $>0$ & $<0$  & warm &  warm \\
IV & $>0$ & $>0$  & KZ &  warm \\
V & $<0$ & $>0$  & KZ & KZ  \\
\hline
\end{tabular}
    \caption{\justifying
In the upper figure we show the plot of the derivative of the collisional integral, evaluated at the Kolmogorov-Zakharov solutions for homogeneous wave action flux and energy flux, as a function of $\beta$. We highlight the region in which the WKE was proved to be well-posed (see \cite{germain2025local}).
Below is a table of the signs of fluxes associated to the two different KZ solutions. We mark with "warm" in contrast to "KZ" the absence of a standard cascade solution.}
    \label{fig:collisional}
\end{figure}

As in most of the literature on the MMT model, from now on in our numerical simulations, we fix $\alpha = 1/2$. We shall not loose generality,  due to the small allowed interval of $\alpha$ in one-dimensional systems. For $\alpha > 1$ the resonant manifold is in fact empty. 
In different regions of the $\beta$-space fluxes have different signs, as shown in fig. \ref{fig:collisional}. 

In the infinite inertial range limit, because the flux of wave action can only be directed towards small wavenumber and the flux of energy only toward high ones, we must then conclude that in regions $II$, $III$ and $IV$ of fig. \ref{fig:collisional} fluxes have the wrong sign, and thus the KZ solutions are not physically relevant. 
It is straightforward to understand why in those specific points the derivative of the collisional integral, and hence the fluxes, changes sign. Using the so-called Zakharov transform, the collisional integral can be rewritten to show explicitly that it possess four zeros. Two of those are always the power laws with exponent $\alpha$ and $0$, corresponding to the two asymptotics of the RJ solution \eqref{RJ}, representing either equidistribution of energy or wave action. Varying $\beta$  does not influence this last two solution exponent, but changes the values of of the KZ ones. When a KZ exponent becomes larger than the lowest RJ one
, the slope of the collisional integral at that point has to change to preserve continuity. When the same solution overcomes the second RJ exponent, the correct slope is restored for the same reason. The four regions of Fig. \ref{fig:collisional} may then be obtained by matching the two KZ solutions to the two RJ ones. In those regions where the flux of the KZ solution has the wrong sign, one expects the stationary state to be close to thermodynamic equilibrium on phenomenological arguments. The common name for such solutions is warm cascades. We will come back on the subject in a later section.
\subsection{WKE for $\alpha = \frac{1}{2}$}
In the next section, we present numerical simulations of the MMT kinetic equation, obtained by varying the $\beta$ parameter and keeping $\alpha=\frac{1}{2}$. 
The collision term in the WKE is a three-dimensional integral.  
To numerically solve it, the two Dirac delta functions are eliminated, reducing the problem to a one dimensional integral. 
One needs to find solutions to the following set of equations that define the resonant manifold:
\begin{gather}
  k+k_1-k_2-k_3 = 0, \\
  \sqrt{|k|}+\sqrt{|k_1|}-\sqrt{|k_2|}-\sqrt{|k_3|} = 0.
\end{gather}
Expressing $k_2$ and $k_1$ in terms of $k$ and $k_3$, and using the properties of the delta function, the WKE reduces to:
\begin{equation}\label{Eq:MMT_redWKE}
\frac{\partial}{\partial t}n_k=4\pi \int_{-\infty}^\infty\frac{\left| kk_1k_2k_3\right|^{\beta/2}}{\Delta_{12}}n_kn_1n_2n_3  \left(\frac{1}{n_k}+\frac{1}{n_1}-\frac{1}{n_2}-\frac{1}{n_3}   \right)dk_3,
\end{equation}
where 
\begin{gather}
    \Delta_{12}=\frac{1}{2}\left| \frac{\text{sign}{(k_2)}}{\sqrt{k_2}}- \frac{\text{sign}{(k_1)}}{\sqrt{k_1}}  \right|, \\
    k_2=k+k_1(k_3)-k_3,\\
    k_1(k_3)=k\, q_1\left(\frac{k_3}{k}\right),
\end{gather} 
with
\begin{equation}
q_1(x) =
\left\{
        \begin{array}{ll}
            -\frac{\left(x+\sqrt{-x}-1\right)^2}{\left(\sqrt{-x}-1\right)^2} & \quad \phantom{-}x \leq -1 \\
            \frac{x \left(x-2 \sqrt{-x}-1\right)}{(x+1)^2} & \quad -1 \leq x \leq 0\\
            \frac{1}{2} \left(\sqrt{8 x^{3/2}-3 x^2-6 x+1}+x-1\right) & \quad \phantom{-}0 \leq x \leq 1\\
            \frac{1}{2} \left(\sqrt{x^2-6 x+8 \sqrt{x}-3}+x-1\right) & \quad \phantom{-}1 \leq x\\
        \end{array}
    \right.
\end{equation}
Note that a different parametrization of the resonant manifold from  \cite{ZAKHAROV2004} has been used for numerical convenience.
\begin{figure}[h!]
\centering
\includegraphics[width=0.48\columnwidth]{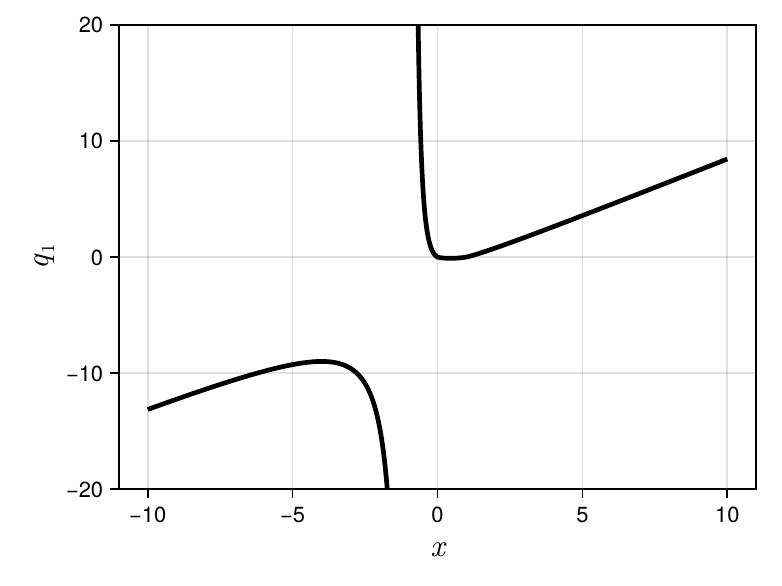}
\caption{
\justifying
\label{fig:resonant} Plot of the resonant manifold for $\alpha = \frac{1}{2}$. $q_1$ identifies the $k_1$ vector normalised by the wave vector $k$ and $x$ represents the $k_3$ vector normalised in the same way.}
\end{figure}
Looking at the plot of this last relationship in Fig. \ref{fig:resonant}, one can clearly see how very different wave vectors interact exchanging conserved quantities in the collisional integral. It is thus very plausible that nonlocal effect play an important role in the statistical dynamics of the system.
\section{Numerical simulations of the Wave Kinetic Equation}\label{sec:num_sim}

We perform numerical simulations of the kinetic equation \eqref{Eq:MMT_redWKE}  with various values of $\beta$ using the WavKinS.jl package, introduced in \cite{Giorgio1} and \cite{Giorgio2}. The WKE is solved on a logarithmic lattice, with wave numbers following a geometric progression, $k_n = k_0 \lambda^n$. The grid parameter $\lambda$ is fixed by the number of grid points $M$ as $\lambda = (k_{max}/k_0)^{1/M}$. 
The KZ solutions are theoretically derived by imposing constant fluxes of wave action and energy. To reproduce this framework numerically, one introduces a forcing term $f(k)$ and a dissipation one $\mathcal{D}(k)$, so that the WKE becomes 
\begin{equation}
    \frac{\partial}{\partial t}n_k = \mathcal{S}\{ n_k\} - f(k) - \mathcal{D}(k),
\end{equation}
where 
\begin{equation}
    \mathcal{D}(k) = \left[(k_{IR}/k)^{-\gamma_{IR}} + (k/k_{UV})^{\gamma_{UV}}\right] n_k,
\end{equation}
$\gamma_{IR}$ and $\gamma_{UV}$ are  real positive parameters governing the dissipation rate, $k_{IR}$ and $k_{UV}$ are the scales at which dissipation takes place. IR and UV refers respectively to large wavelengths (Infrared) and small ones (Ultraviolet). The forcing has a Gaussian shape to manipulate freely its support.
Starting with almost zero wave action and energy, we let the system evolve under the effects of the forcing, properly placed as to induce a wave action cascade or an energy one. What we show in the following plots are the long time stationary states that we observed. The stationarity of such states is reflected by the flat region of the energy and wave action fluxes, normalized against the respective injection rates, shown in the provided insets. We focus on regions III and IV, as numerical simulations become increasingly expensive for larger values of $|\beta|$. 

\begin{figure}[h!]
    \centering
    \begin{subfigure}[b]{0.48\columnwidth}
        \includegraphics[width=\linewidth]{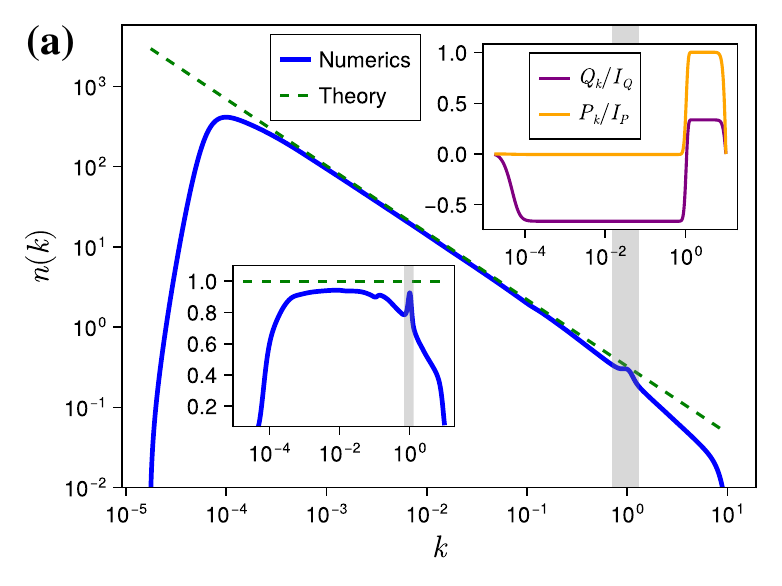}
        \phantomcaption
        \label{fig:inverse0}
    \end{subfigure}
    \begin{subfigure}[b]{0.48\columnwidth}
        \includegraphics[width=\linewidth]{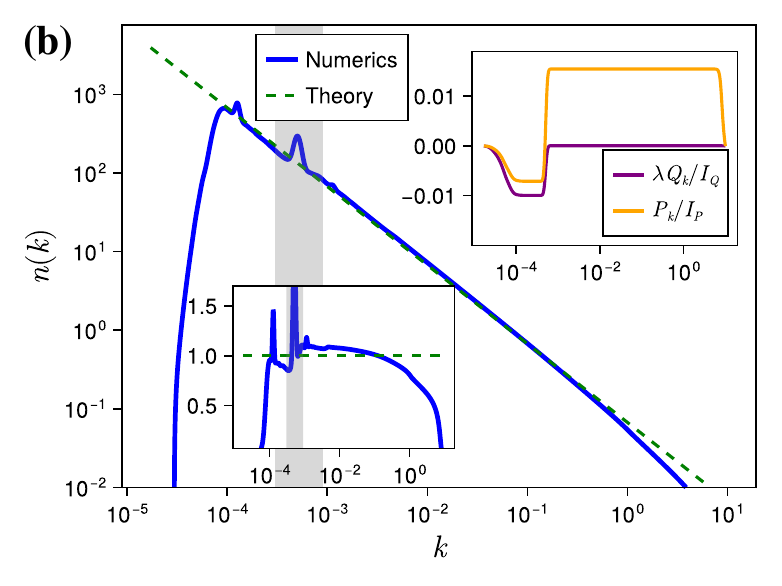}
        \phantomcaption
        \label{fig:direct0}
    \end{subfigure}
    \caption{
        \justifying
        \textbf{(a)} Log-log plot of the stationary state of a simulation with $\beta = 0$ and $M = 5000$ points. The lower inset shows the same data compensated by the KZ state in a lin-log plot. The upper inset shows a lin-log plot of the wave action and energy fluxes, normalised by the corresponding injection rates.
        \textbf{(b)} Similar to \textbf{(a)} but for the direct cascade. In both pictures the gray area represent the region where the forcing is noticeably different than 0.
    }
    \label{fig:combined}
\end{figure}

The first simulation we present is performed with $\beta=0$. We run two simulations, one with the forcing at low $k$ and the other at high $k$. The remarkable agreement with the KZ theoretical prediction, across four decades, can be observed both in fig. \ref{fig:inverse0} and fig. \ref{fig:direct0}. Notice how in this last fugure  the cascade is not completely developed for high $k$. The reason is that the direct cascade for $\beta = 0$ possesses infinite capacity, meaning that the total energy contained in the stationary state distribution diverges as $k_{UV} \rightarrow \infty$. Consequently, the higher the wave vector, the longer the time required to saturate its energy content. 

\begin{figure}[h!]
\centering
\includegraphics[width=0.48\columnwidth]{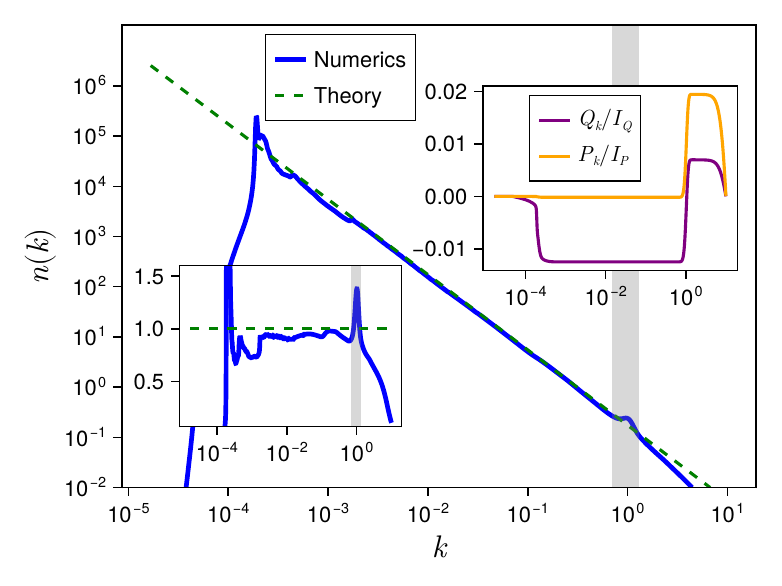}
\caption{
\justifying
\label{fig:inverse1} Log-log plot of the late time stationary state of a simulation with $\beta = 1$ and $M = 2000$. In gray it is represented the region where the forcing is noticeably different than 0.
In the lower inset the same data are weighted against the KZ state in a lin-log plot.
In the upper inset we show a lin-log plot of the wave action and energy fluxes in the stationary state, weighted against the corresponding injection ranges. The gray area represent the region where the forcing is noticeably different than 0.}
\end{figure}

It has recently been proved that the kinetic equation is well posed in the range $\beta \in (-\frac{1}{2},0)$, \cite{germain2025local}: 
here, we numerically investigate the behavior of the KZ solutions in a regime where rigorous mathematical results are currently unavailable. For higher $\beta$, given that the fluxes maintain the proper signs, and the locality window extends far above $\beta=0$, on physical grounds one would not expect qualitative changes in the behavior of the solution of the kinetic equation. The late time stationary state in an inverse cascade with $\beta=1$ and $M = 2000$ may be observed in fig. \ref{fig:inverse1}. Again, the agreement with the KZ prediction and the existence of a stationary state are clear. 

\begin{figure}[h!]
    \centering
    \begin{subfigure}[b]{0.48\columnwidth}
        \includegraphics[width=\linewidth]{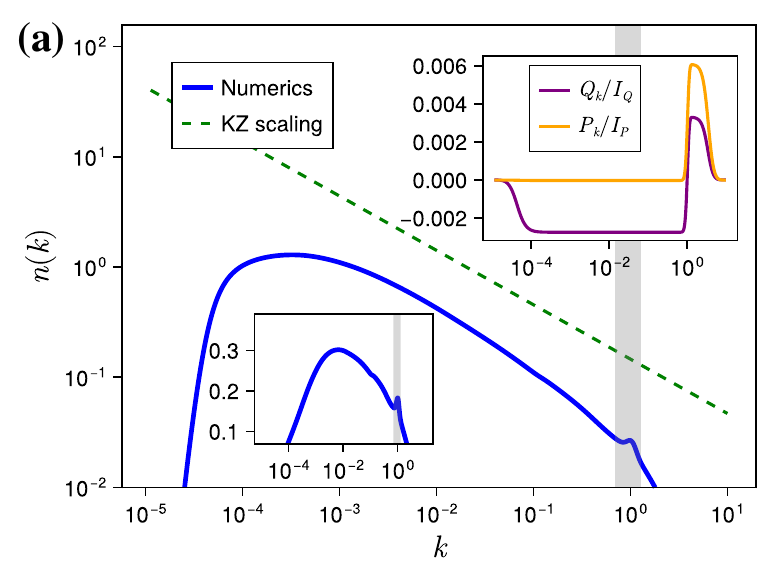}
        \phantomcaption
        \label{fig:inverse-51}
    \end{subfigure}
    \begin{subfigure}[b]{0.48\columnwidth}
        \includegraphics[width=\linewidth]{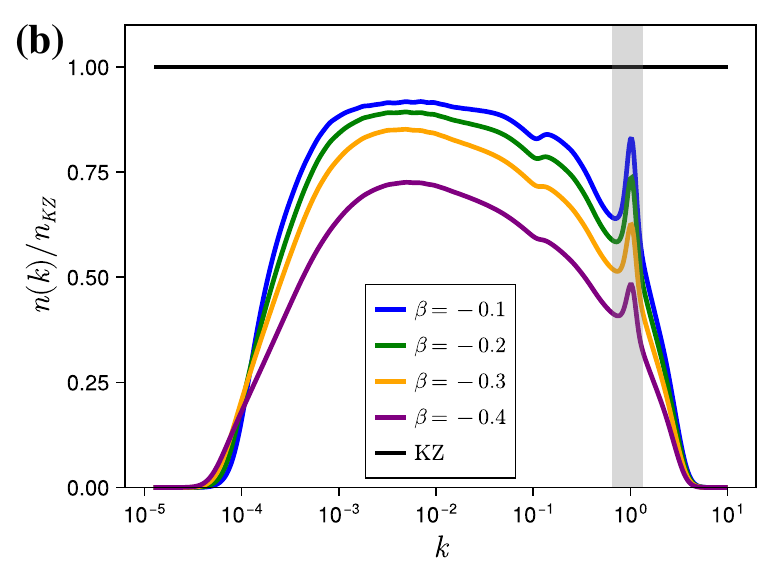}
        \phantomcaption
        \label{fig:phase_transition}
    \end{subfigure}
    \caption{
        \justifying
        \textbf{(a)} Log-log plot of the stationary state of a simulation with $\beta = -0.51$ compared to $k^{-\gamma_q}$. 
        In the lower inset the same data are compensated by $k^{-\gamma_q}$ in a lin-log plot. Being below the threshold $\beta = 1/2$, we cannot refer to KZ solutions in this case, as their coefficient is negative.
        In the upper inset we show a lin-log plot of the wave action and energy fluxes in the stationary state, normalised by the corresponding injection ranges. The gray area represents the region where the forcing is noticeably different than 0.
        \textbf{(b)} Various simulations performed at different $\beta$s and with $M = 500$, weighted by the respective KZ solutions.
    }
    \label{fig:combined}
\end{figure}

So far, the behavior of the Kinetic Equation is in very good agreement with the predictions of Zakharov’s analytical theory. It is now interesting to explore the region of $\beta$-space where one of the fluxes changes sign. To this end, we perform a numerical simulation with $\beta = -0.51$, just at the boundary between regions $IV$ and $V$ in fig. \ref{fig:collisional}, so that the expected solution has the wrong flux sign for the inverse cascade. 
What can be clearly observed in fig. \ref{fig:inverse-51} is that a stationary state exists but not in the fashion of a power law. Here, there is no effective mechanism of wave action transport. Instead, wave action is transferred in a much less efficient way, as it can be seen by the wave action flux, at least one order of magnitude smaller than in the other cases presented. 
We were not able to observe a sharp transition by simulating the system slightly above and below the threshold $\beta=-1/2$; the change in sign of the flux is predicted by a theory which assumes an infinite inertial range, while necessarily the one observed in numerical simulations is bound to be finite. To observe the transition from the KZ state to the novel one, we performed a number of lower resolution simulations, at various values of $\beta$, whose stationary states can be observed in fig. \ref{fig:phase_transition}. Then it becomes manifest that lowering the $\beta$ parameter results in a gradual departure from a power law cascade. We stress that, for $\beta < -1/2$,  the flux of the KZ solutions is in the opposite direction than expected. The direct consequence is a change of sign in the KZ constants defined in eq. \eqref{KZ P 2} and \eqref{KZ Q 2}. The KZ power laws are still zeroes of the collisional integral, but from the physical point of view are unacceptable, as the wave action density $n_k$ is defined as a positive quantity. One can expect that,  given the forcing's injection of conserved quantities, nonlinear interactions still cooperate to transfer energy and wave action across scales, albeit in a less effective manner. It can be argued that the system will stabilize towards an approximate thermodynamical equilibrium, where $T$ and $N$ are fixed by the fluxes ($T \propto P_0^{1/3}$). Those kind of solutions have been named warm cascades in the literature. The reader is directed to \cite{Nazarenko2011} for a thorough discussion of the subject.  

It is well known that Wave Turbulence Theory predicts the average behavior of the microscopic differential equations in a restricted number of situations. Numerous deviations are observed numerically in the MMT model, see \cite{ZAKHAROV2004}. Wether they arise due to the formation of coherent structures or result from strong pumping of conserved quantities, there is not to this date a satisfactory theoretical framework able to resolve the shortcomings of Wave Turbulence. A possible direction consists in deriving the WKE considering also higher order perturbative contributions. In the next section, we study the consequences of such an expansion for the MMT model, using recently introduced techniques.

\section{Kinetic Equation beyond the leading order}\label{sec:beyond}
\subsection{Next to Leading order Wave Kinetic Equation}
Wave turbulence is bound to be effective only at weak nonlinearity. The derivation of the Wave Kinetic Equation truncates every order of perturbation theory except the first nontrivial one. At growing perturbative orders calculations become quickly rather involved. For this reason diagrammatic methods were introduced, for example in \cite{zakharov1975statistical} and \cite{Gurarie:1994ut}. More recently,  the diagrammatic tools of Quantum Field Theory were applied to Wave Turbulence in \cite{rosenhaus2024wave}, \cite{Rosenhaus2023}, \cite{Rosenhaus:2023sik}, \cite{Rosenhaus:2023pdj}. 

Rosenhaus and Smolkin, \cite{Rosenhaus2023}, mapped the problem of truncating the infinite set of differential equations for correlators into that one of perturbatively evaluating expectation values of a statistical theory defined by a path integral.
By evaluating the Feynman diagrams of the first two perturbative orders for the two- and four-point correlation function, they obtained a WKE at the next-to-leading order in the general case of four wave interactions and general dimension. The general expression they obtained is reproduced in the Appendix \label{App:NLOWKE}, eq.~\eqref{NLO_general}.  Following their work, we find for the one dimensional MMT model with generic $\beta$ and $\alpha = 1/2$ the following next to leading order Wave Kinetic equation:
\begin{gather}
        \frac{\partial}{\partial t}n_k(t) = 4\pi \int_{-\infty}^\infty dk_1dk_2dk_3\left[|kk_1k_2k_3|^{\beta/2} + \sigma(k,k_1,k_2,k_3)\right]
         n_kn_1n_2n_3\notag\\
        \times   \left(\frac{1}{n_k} +\frac{1}{n_1}-
        \frac{1}{n_2}-\frac{1}{n_3}\right)\delta\left(\Delta k^{k1}_{23}\right)\delta\left(\Delta\omega^{k1}_{23}\right), 
        \label{wke_nlo}
\end{gather}
    where 
\begin{multline}
        \sigma(k,k_1,k_2,k_3) =  4 |kk_1k_2k_3|^{\beta/2}\int_{-\infty}^\infty dk_4dk_5n_4n_5|k_4k_5|^{\beta/2} \\
        \times \left[\left( \frac{1}{n_4}+\frac{1}{n_5} \right) 
        \frac{\delta(\Delta k^{45}_{23})}{\sqrt{|k_2|}+\sqrt{|k_3|}-\sqrt{|k_4|}-\sqrt{|k_5|}} \right.
        \\ + 4\left( \frac{1}{n_4}-\frac{1}{n_5} \right) 
        \left.\frac{\delta(\Delta k^{35}_{14})}{\sqrt{|k_3|}+\sqrt{|k_5|}-\sqrt{|k_1|}-\sqrt{|k_4|}}\right].
        \label{sigma}
\end{multline}

\begin{figure}[h!]
\centering
\includegraphics[width=0.3\columnwidth]{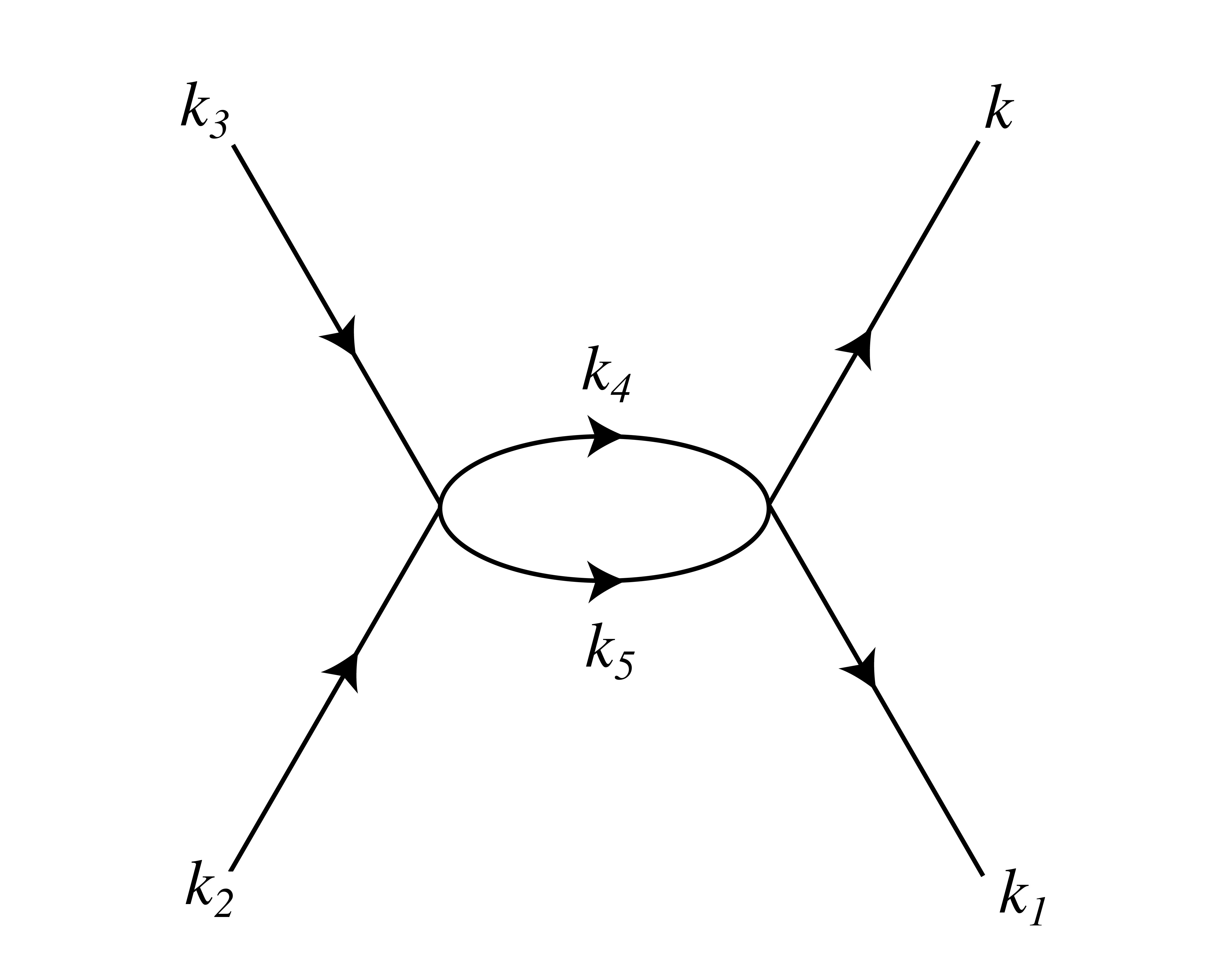} \\
\includegraphics[width=0.3\columnwidth]{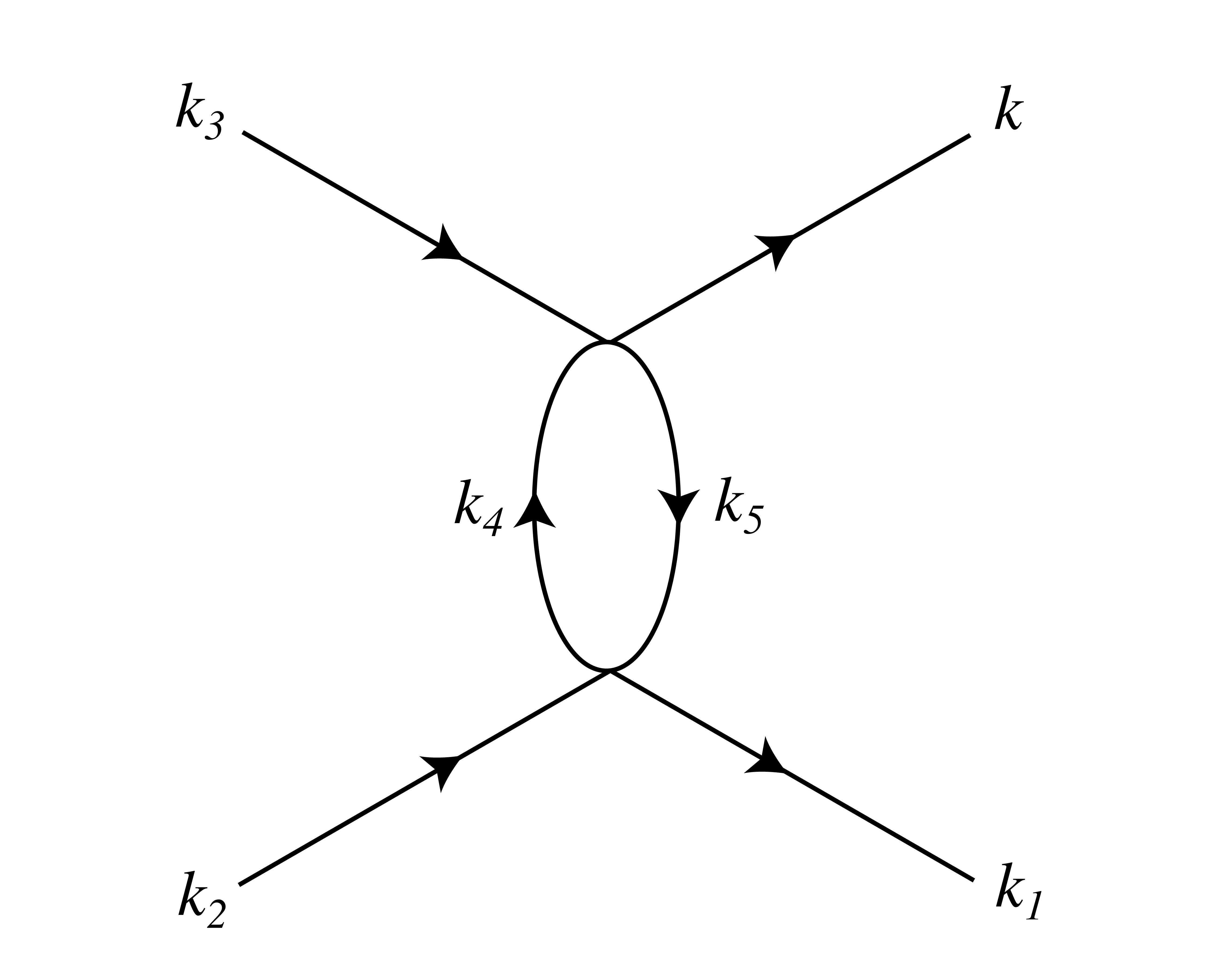}
\includegraphics[width=0.3\columnwidth]{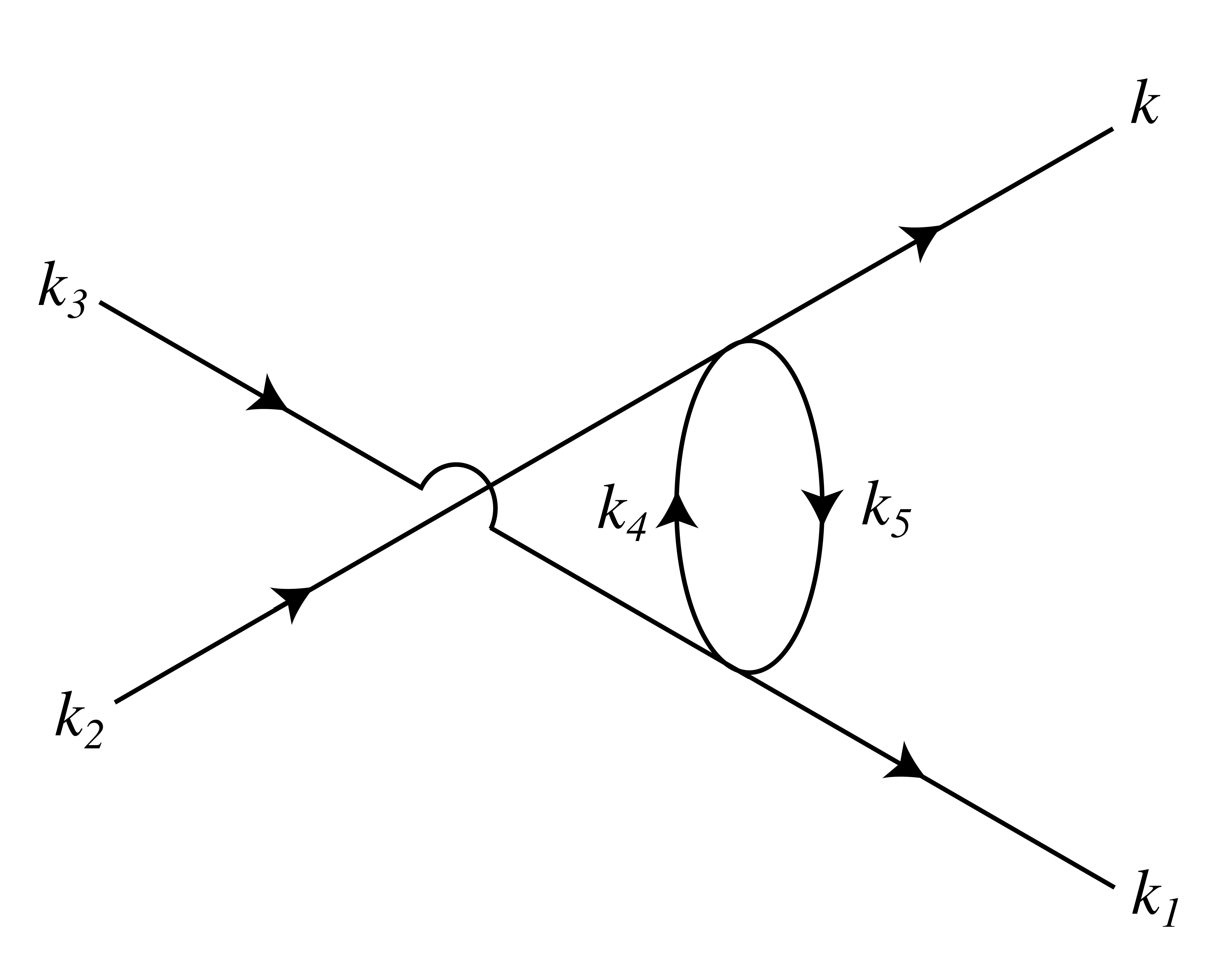}
\caption{
\justifying
\label{fig:diagrams}Most relevant diagrams for the derivation of the next to leading order wave kinetic equation.}
\end{figure}

This form of the equation is the result of extensive manipulations. It is possible to show that on the resonant manifold
\begin{equation*}
    \sigma(k,k_1,k_2,k_3) =  
    \sigma(k_1,k,k_2,k_3)  = 
    \sigma(k_2,k_3,k,k_1) .
\end{equation*}
An important consequence of this symmetries is that linear energy and wave action are still conserved, even at higher perturbative order. As in many areas of physics, convoluted perturbative expansions are described order by order through diagrammatic techniques. Every mathematical expression is associated to a pictorial representation, from which it can be easily recovered. The bookkeeping of the perturbative calculations is then performed by manipulating directly the diagrams following a specific setting of rules, and only at the end they are transformed back to the explicit form to perform actual calculations. This type of diagrammatic techniques are generally known as Feynman diagrams. In the case of eq. \eqref{sigma} the first contribution originates from the upper diagram in fig. \ref{fig:diagrams}, whereas the second term corresponds to the combined expressions encoded in the two lower diagrams. Without entering in the details, we highlight that the perturbative order is equal to the number of vertices in the diagram and that each line originating and ending inside the diagram corresponds to an additional integration. To each line a variable is associated, indicating the argument of corresponding mathematical expression. The reader interested in the details of the expansions, and how to extract the kinetic equation from it, is directed to \cite{Rosenhaus2023}.   

As highlighted in \cite{Rosenhaus:2023pdj}, very often the collisional integral of the next to leading order equation is characterized by an empty locality window, meaning that no power law solution choice for $n_k$ will make the integral converge both at high and low $k$. Consequently, in general, the hypothetical out of equilibrium stationary state will be dependent on the position in $k$-space of the forcing. 
Those divergencies are however not too much worrisome, as the presence of forcing and dissipation produce automatically cutoffs both at high and low wave number. Also, it can be argued that all classical systems possess natural cutoffs, induced by finite size at low $k$ and by the smallest scale at which the original differential equation is valid at high $k$. 

Our study unveiled a second kind of divergence present in this equation, arguably of a worse nature.  
Let us look at the first term in \eqref{sigma}
    \begin{equation}
        \frac{\delta(\Delta k_{23}^{45})}{\sqrt{k_2}+\sqrt{k_3}-\sqrt{k_4}-\sqrt{k_5}},
    \end{equation} 
    Using fig. \ref{fig:diagrams}, we can interpret the $\sigma$ term as the exchange of virtual waves among different modes.
    $k$ and $k_1$ produce two virtual waves ($k_4$ and $k_5$) that rescatter into $k_2$ and $k_3$.
    Being virtual, they do not need to be resonant, but the closer to resonant they are, the bigger their contribution to the process, as the denominator suggests.   
    We recast it by using the delta function in the numerator, removing in the process the $k_5$ integration, and rename $k_4=q$. Remember that $\sigma$ arguments $k\dots k_3$ have to be part of the resonant manifold, hereby called external, implying that
    \begin{equation}
        \begin{aligned}
      D_1 &= \sqrt{k_2} + \sqrt{k_3} -\sqrt{q} - \sqrt{k_2 + k_3 - q}\\
       &= \sqrt{k} + \sqrt{k_1} -\sqrt{q} - \sqrt{k + k_1 - q}.
        \end{aligned}
    \end{equation}
    It is easy to see that this expression has four zeroes corresponding to $q$ equal to any of the external wave numbers. 
    As long as $k \neq k_1$ or $k_2 \neq k_3$ the zeroes are singles. For example, let's look at $q = k$; we rename $p = q-k$ and expand around $p=0$ to obtain
    \begin{equation}
      \sqrt{|k|} + \sqrt{|k_1|} -\sqrt{|p+k|} - \sqrt{|k_1 - p|} \underset{p \rightarrow 0}{\approx} 
      \frac{1}{2} p \left(\frac{\text{sign}(k)}{\sqrt{|k|}} - \frac{\text{sign}(k_1)}{\sqrt{|k_1|}}\right).
    \end{equation}
    In this case, under the assumption that $n_k$ is continuous, $\int dp/D_1$ would behave close to $p=0$ as $\int_{-\epsilon}^{+\epsilon}dp/p$, whose Cauchy principal value is 0 (as all other factor in the integrals are regular on that point). Remember that we are looking at $k_4 \rightarrow k$, and all other terms present in the integral do not contribute to the diverging behavior. \\
    However, if $k_1 = k$ we have to move to second order in the expansion, finding 
    \begin{equation}
      2\sqrt{|k|} -\sqrt{|p+k|} - \sqrt{|k - p|} \underset{p \rightarrow 0}{\approx} \frac{1}{2}|k|^{-\frac{3}{2}} p^2.
    \end{equation}
    This is a problem, as the principal value of $\int_{-\epsilon}^{+\epsilon}\frac{1}{p^2}$ diverges. The same line of reasoning can be easily followed in the case of the second term of eq. \eqref{sigma}.
    We must remember that the $\sigma$ term is multiplied to $\frac{1}{n_k} + \frac{1}{n_1} -\frac{1}{n_2} -\frac{1}{n_3}$ in the full kinetic equation. This means that, if in the external manifold $k = k_1$ implies $k_2 = k_3$, the divergence in $\sigma$ is healed by multiplication by zero in the kinetic equation. If instead there exist nontrivial points where $k=k_1$ corresponds to $k_2 \neq k_3$, we are looking at a true divergence.
    We set $k_1=k$ in the equations defining the external resonant manifold, the system to solve is thus
    \begin{gather}
            2k = k_2 + k_3 \notag\\
            2\sqrt{|k|} = \sqrt{|k_2|} + \sqrt{|k_3|}
        \label{resmanifold}
    \end{gather}
    Assuming $k>0$,  we find the trivial solution ($k_2 = k_3 = k$) and two nontrivial ones,
    $k_2 = -\frac{k}{4} \hspace{1mm} \& \hspace{1mm}k_3 = \frac{9k}{4} $ and $k_3 = -\frac{k}{4} \hspace{1mm}\& \hspace{1mm}k_2 = \frac{9k}{4} $ obtained through direct substitution of the first equation into the second. The sign of the integrand is the same in both cases, not allowing for possible cancellations.
    The second denominator of \eqref{sigma} does not present any of those incurable divergencies. The validity of the perturbative expansion is to be questioned when this kind of divergences appear, and it is natural to ask if other systems suffer from them as well.
    
    One may wonder if with a different dispersion relation, or in higher dimension, the situation changes. We argue this is not the case. 
    In the case of generic $\alpha$ and higher number of spatial dimensions in eq. \eqref{NLO_general}, double zeroes of $\Delta \omega_{45}^{23}$ are still incurable divergencies of the inner integrals. Given that they are located in the middle of the inertial range, no terms coming from angular integration can balance them. It is then enough to study the system 
\begin{gather}
            2\vec{k} = \vec{k}_2 + \vec{k}_3\notag \\
            2{|\vec{k}|}^\alpha = {|\vec{k}_2|}^\alpha + {|\vec{k}_3|}^\alpha
        \label{resmanifold2}
\end{gather}
Defining $t = k_2 / k_3$ and using $|\vec{k}_2||\vec{k}_3|\cos\theta = \vec{k}_2 \cdot \vec{k}_3$, after substituting the first equation of \eqref{resmanifold2} into the second one, we obtain for any dimension $d$, the following equation 
\begin{equation}
    \Lambda(t) \equiv \frac{2^{2-2/\alpha}(1+t^\alpha)^{2/\alpha}-1-t^2}{2t} = \cos\theta.
    \label{divergencies_eq}  
\end{equation}
\begin{figure}[h!]
\centering
\includegraphics[width=0.45\columnwidth]{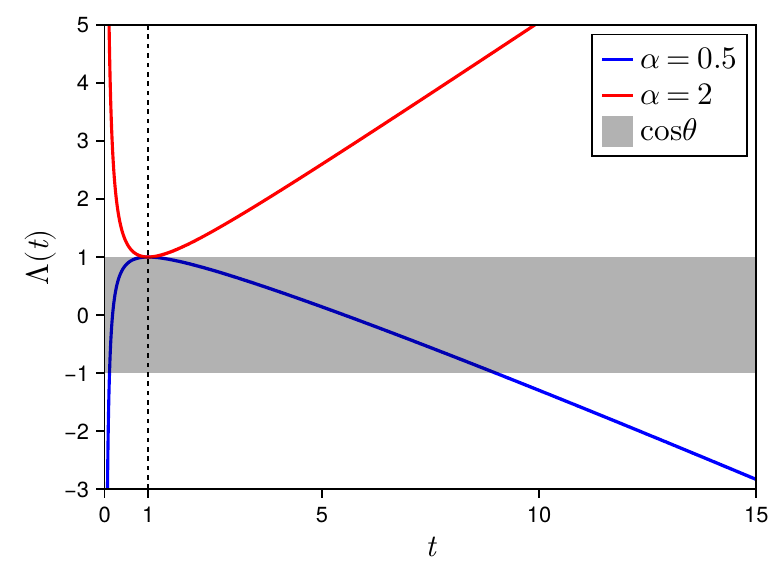}
\caption{\label{divergencies}\justifying L.h.s. of equation \eqref{divergencies_eq} for two different values of $\alpha$, corresponding to surface water waves and the Nonlinear Schrodinger equation. The gray area represents the possible values that the r.h.s. can assume.} 
\end{figure}
We note that non-trivial solutions exist if $\Lambda(t)$ takes values in $(-1,1)$. As $\Lambda(1)=1$ and $\Lambda^{''}(1) = \alpha - 1$, it is then concave for $\alpha<1$ and solutions always exits, implying the existence of divergences in the next-to-meaning order WKE. Conversely, for $\alpha>1$ we can show numerically that there are no solutions and therefore such divergences do not exist. In fig.~\ref{divergencies}, we plot the l.h.s. of equation \eqref{divergencies_eq} for $\alpha=1/2$ and $\alpha=2$, which correspond to the two well known examples of surface water waves in infinite water depth and the NLS equation, respectively.  For $\alpha=2$, we notice that the red line enters the gray area only for $t = 1$, implying $\cos\theta = 1$, and thus showing that the only solution to the above system in that case is the trivial solution $\vec{k}_2 = \vec{k}_3$, leading to a curable divergence. That is not the case for the blue line that presents all kinds of possible resonances. This finding shows that whereas the perturbative expansion might be well controlled for the NLS equation, for gravity waves it presents strong divergences, and therefore some other approaches need to be found.

\subsection{Beyond perturbation theory}

Numerous efforts have been devoted during the years to the aim of overcoming the intrinsic limitations of perturbative theory in wave turbulence. For example, the Direct Interaction Approximation (DIA) developed by Kraichnan (\cite{Kraichnan1959}) was used to extract re-summed equations, in the case of the NLS equation in \cite{DuBois1981} and for elastic plates in \cite{Pavez2023}.
The closed set of expressions is obtained by substituting in the perturbative expansion for the two point function the unknown exact propagators to the free theory ones.
The equations derived this way are unwieldy, but seem to be non pathological, and reduce to the WKE of the respective systems in the proper limits. More numerical and theoretical work is required to understand if the closure is even meaningful, and to extract predictions from the theory.\\
In a recent work, \cite{largeN}, the perturbative expansion for a generic four wave system was re-summed employing the large N expansion. Considering a system of N interacting identical components, the upper diagram of fig. \ref{fig:diagrams} and its higher perturbative equivalents are found to be sub-leading in the large N limit. Given that the expression represented in them is the one associated to the incurable divergencies we discussed, the resulting kinetic equation does not suffer anymore from them. Of course, this consideration requires a careful analysis of the weakly nonlinear and large N limiting procedure. The possibility of applying it effectively for low values of N is still under discussion, and requires further investigation.       


\section{Conclusions}\label{sec:conclusions}

In this work, we numerically confirmed the predictions of Wave Turbulence theory for different parameters of the MMT model. Kolmogorov-Zakharov solutions are observed to exist and appear to be stable for the prototypical case of $\beta = 0$ (NLS) and surface water waves dispersion relation. KZ solutions are observed also for  higher values of $\beta$, where the kinetic equation is not yet proved to be well posed. 

We also investigated the region where no cascade solutions are expected, because of a wrong sign in their flux. Surprisingly, we observe an apparently stable stationary state mimicking an inverse cascade, but with much less effective wave action transport properties. By using multiple numerical simulations with similar nonlinear parameters, we observe the transition to this new state, showing that indeed there exists a drastically different stationary regime. Further work is necessary to characterize the transition and the novel state, theoretically and numerically. Such work could also help confirming wether this new stationary state is indeed a warm cascade or not.  

We concluded by analyzing the next to leading order kinetic equation for the model at hand, uncovering incurable divergencies that hinders the applicability of the improved kinetic theory. We extend our argument to higher dimensional four wave kinetic equations whose dispersion relations are concave power laws of the wave vector's absolute value.
Luckily, other avenues are possible in the same spirit of overcoming perturbative limitations of wave turbulence, like the Direct Interaction Approximation and the large N limit, that both appears to be free of the divergencies we studied.

\enlargethispage{20pt}

\subsection*{Acknowledgments}
This work was supported by the Simons Foundation (USA) under Awards No. 652354 and  No. 651471 (Wave Turbulence). M.O. was additionally supported by INFN through the MMNLP and FIELDTUR projects. The authors thank V. Rosenhaus for valuable and stimulating discussions.


\appendix
\section{Dimensionless collisional integral} \label{section:Aappendix}
The dimensionless collisional integral, after defining $q_i = \frac{k_i}{k}$, reads as 
\begin{multline}
    \mathcal{I}(\nu) = \int_{-\infty}^\infty dq_1 dq_2 dq_3 \left( 1 + |q_1|^{-\nu} - |q_2|^{-\nu} - |q_3|^{-\nu}\right) \times \\
    \left( q_1 q_2 q_3\right)^{\frac{\beta}{2} -\gamma} \delta(1 + |q_1|^\alpha  - |q_2|^\alpha - |q_3|^\alpha)
     \delta(1 + q_1 - q_2 - q_3).
\end{multline}

\section{General Next to Leading Order wave kinetic equation}
We present, without derivation, the full next to leading order kinetic equation for a $D$ dimensional four wave system with complex interaction term, obtained in \cite{Rosenhaus2023}.
\begin{multline}
        \frac{\partial}{\partial t}n_k = 4\pi \int \prod_{j=1}^{3} d^Dk_j|T_{k123}|^2n_kn_1n_2n_3\left(\frac{1}{n_k} +\frac{1}{n_1}-
        \frac{1}{n_2}-\frac{1}{n_3}\right)\delta(\Delta \omega^{k1}_{23})\delta^{k1}_{23} \\
          + 16\pi\int \prod_{j=1}^{5} d^Dk_j \delta(\Delta \omega^{k1}_{23}) n_k n_1n_2n_3n_4n_5 \times \\
          \left\{\dfrac{\text{Re}(T_{23k1}T_{k145}T_{4523})}{\Delta \omega^{23}_{45}}\delta^{56}_{34}\delta^{12}_{56}
          \left(\frac{1}{n_k} +\frac{1}{n_1} -\frac{1}{n_2} - \frac{1}{n_3}\right)\left(\frac{1}{n_4}+\frac{1}{n_5}\right)  \right. \\
          + \frac{3}{2}\pi\text{Im}(T_{23k1}T_{k145}T_{4523})\delta(\Delta \omega^{23}_{45})\left(\frac{1}{n_k}+\frac{1}{n_1}\right)\left(\frac{1}{n_2}
          +\frac{1}{n_3}\right) \\
            + 4\left[\dfrac{\text{Re}(T_{23k1}T_{k524}T_{1435})}{\Delta \omega^{35}_{14}}\delta^{25}_{46}\delta^{61}_{35}
            \left(\frac{1}{n_k} +\frac{1}{n_1} -\frac{1}{n_2} - \frac{1}{n_3}\right)\left(\frac{1}{n_4}-\frac{1}{n_5}\right)\right. \\
            + \left. \left. \frac{3}{2}\pi\text{Im}(T_{23k1}T_{k524}T_{1435})\delta(\Delta \omega^{35}_{14})\left(\frac{1}{n_k}-
            \frac{1}{n_2}\right)\left(\frac{1}{n_1}-\frac{1}{n_3}\right)  \right]\right\},
            \label{NLO_general}
\end{multline}
where $\delta^{ij}_{lm} = \delta^D (k_i + k_j - k_l - k_m)$.


%

\end{document}